# Magnetically Tunable Chiral Phonon Polaritons with Magneto-optical Bound States in the Continuum

Yu Sun[1], Jue Li[1], Wei Li[1], Bo Li[1,2], Qinghua Song[1, *], and Mengyao Li[1, *]

[1] Institute of Materials Research, Tsinghua Shenzhen International Graduate School, Tsinghua University, Shenzhen 518055, China
[2] Suzhou Laboratory, Suzhou 215000, China

**Abstract**

Chiral phonon-polaritonic states are of interest for handedness-dependent light–matter interactions, yet their realization and magnetic control remain challenging, while direct magneto-optical tunability of phonon-polaritonic media is limited. Here, we propose a hybrid platform in which an hBN phonon polariton couples to a chiral bound state in the continuum supported by a magneto-optical photonic crystal, enabling strong and selective photonic coupling. The interaction gives rise to pronounced mode splitting and the formation of hybrid states, and their modal composition is quantified by phonon-proportion analysis and described by a coupling theory. Importantly, the hybridization can be controlled by magnetic bias through the magneto-optical response of the photonic component, providing control over the modal composition and spectral response. In addition, the hybrid states exhibit handedness-selective absorption under circularly polarized excitation. This work offers a feasible route toward magnetically tunable chiral phonon-polaritonic devices and hybrid polaritonic functionalities.

## 1. Introduction

Polaritons as quasi-particles induced by strong coupling between light and matter interaction, help to break through the existing diffraction limit and achieve micro-nano scale optical manipulation[1-8]. Phonon polariton is a hybrid mode formed by the coupling of the electromagnetic field and optical phonons in polar media. Its physical origin lies in the oscillating dipole response generated by the coupling of the polar lattice vibrations within the phonon material with the electromagnetic field, enabling deep sub-wavelength light field compression in the mid-infrared to terahertz wavelength range[9-15]. Leveraging their strong localized fields and propagable polariton modes, these systems have been widely applied in scenarios such as waveguide transmission[16,17], wavefront control[18], and thermal management[19,20].

Realization of chiral phonon polaritons is expected to play an irreplaceable role in chirality detection and information transmission in the mid-infrared spectrum. However, there are still problems in chiral phonon polaritons that remain unsolved. For instance, conventional phonon-polaritonic systems are still constrained by limited spectral tunability and inefficient free-space excitation due to in-plane momentum mismatch. To solve this problem, magnetic fields offer an alternative manipulation method for chirality inversion in a single structure while directly breaking time-reversal symmetry and coupling chirality to valley, spin, and magnetic degrees of freedom[21-25]. However, since typical phonon materials such as hBN have their infrared optics mainly dominated by polar lattice vibrations and lack strong magnetic field-sensitive electronic resonance channels, their intrinsic phonon responses usually show only very weak magnetic field dependence. These issues greatly limit the application of phonon polaritons in fields such as ultra-sensitive detection.

Therefore, introducing bound states in the continuum (BICs) into the phonon-polariton platform can provide new design freedom to overcome these obstacles. BIC is a peculiar electromagnetic mode embedded in the radiation continuum yet remains nonradiative[26-28]. Since BIC can theoretically achieve an infinitely large Q factor, indicating extremely high energy localization characteristics, it can significantly enhance the interaction between light and matter[29-35]. Therefore, BIC phonon polaritons are particularly promising for ultrasensitive sensing[36], thermal emission[37] and strong-coupling-based active polaritonic devices[38-40]. Recently, some research teams have investigated the effect of magnetic fields on the characteristics of BIC in photonic crystals[41-43], offering a promising photonic route for magnetic-responsive chiral photonic modes. However, their coupling to phonon polaritons for realizing magnetically controllable chiral hybrid states, and whether the hybrid mode is truly tunable, remains underexplored.

In this work, we propose a hybrid platform by applying magneto-optical BIC photonic crystals onto hBN phonons and achieve magnetically tunable chiral phonon polaritons. First, we constructed the underlying magneto-optical BIC photonic crystal. When the photonic crystal is exposed to an external magnetic field, the far-field polarization of BIC shows chiral differences. Subsequently, by coupling this photonic crystal structure with a thin layer of hBN, which is originally not responsive to the magnetic field, phonon polaritons were successfully obtained. The introduction of an external magnetic field can effectively alter the proportion of phonons to photons in the hBN phonon polariton, which enables phonon proportion tuning by magnetic fields. Moreover, phonon polaritons exhibit magnetic circular dichroism selectively to the excitation of external magnetic fields. This design and theoretical framework provide new theoretical guidance and application interests for the light-matter interactions, and offer a new research direction for the application of chiral polaritons.

## 2. Results and discussion

The chiral phonon polariton platform consists of the magneto-optical photonic crystals placed on the $SiO_2$ substrate, and an additional layer of hBN thin film is applied on the topmost layer to study polaritons. The schematic diagram of the geometric model used in this work is shown in Figures 1a and b. The bottom side length of this magneto-optical photonic crystal is $L = 3000\ nm$, the thickness $h = 3000\ nm$, the radius of the cylindrical air holes inside the cubic photonic crystal is $r = 600\ nm$, and the side length of each unit cell is $a = 4100\ nm$. An asymmetric parameter $d$ is introduced in order to control the effective coupling of the BIC mode with the hBN phonon, which represents the distance from the cylindrical air hole to the center of the cube. The refractive index of this magneto-optical photonic crystal is shown as follows[44,45]:

$$\hat{n} = \begin{pmatrix} n_0 & i\delta & 0 \\ -i\delta & n_0 & 0 \\ 0 & 0 & n_0 \end{pmatrix} \quad (1)$$

In our model, $n_0$ is set to 2.83, $\delta$ represents the magnitude of the external magnetic field. When the magnetic field is incident perpendicularly on the photonic crystal along the $z$ direction, its refractive matrix is affected. When $\delta = 0$, it indicates that there is no external magnetic field. The relative permeability of this material is $\mu_r = 1$.

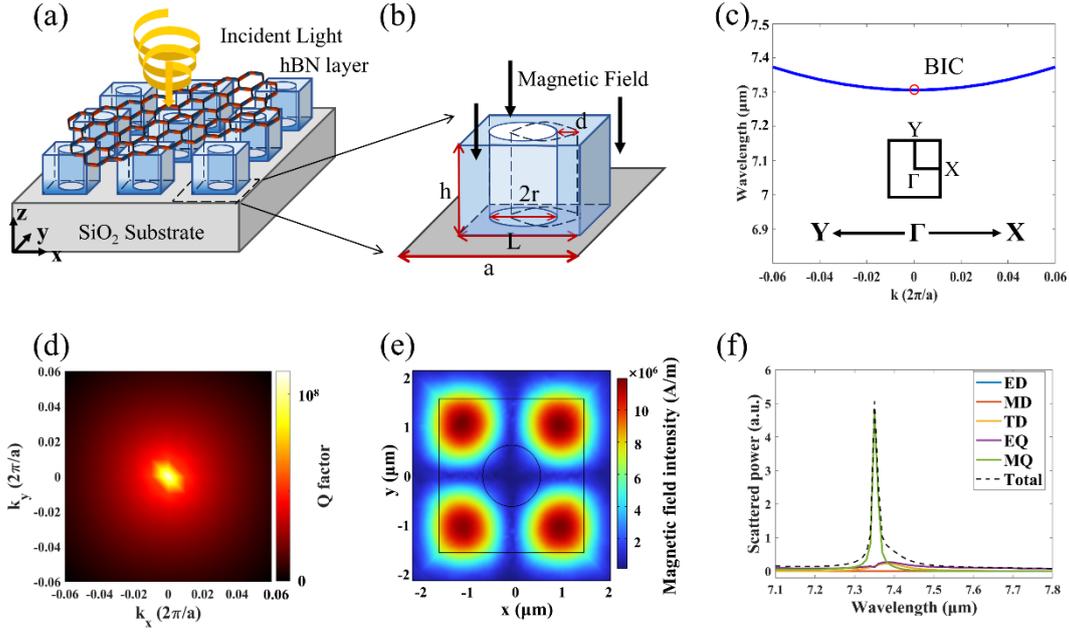

**Figure 1. Structural design of the phonon polariton system and BIC properties of the underlying photonic crystal.** (a) The sketch of magneto-optical photonic crystal slabs with the SiO$_2$ substrate, and a thin layer hBN deposited on the top. (b) Schematic diagram of the geometric parameters of the unit cell. The structure of the unit cell of magneto-optical photonic crystals includes parameters such as bottom side length $L$, thickness $h$, radius of cylindrical air holes $r$, periodic constant $a$ and asymmetric parameter $d$. (c) Band diagram of the magneto-optical photonic crystal without SiO$_2$ substrate and $d = 0\ nm$. (d) The quality factor Q distribution of the corresponding BIC energy band. The Q value at $\Gamma$ point exceeds $10^8$. (e) Distribution of the longitudinal magnetic field $H_z$ in the xOy plane. (f) The multipole decomposition results of the BIC mode. The scattered power of the multipole components in the BIC is calculated based on their respective multipole moments, including electric dipole (ED), magnetic dipole (MD), toroidal dipole (TD), electric quadrupole (EQ) and magnetic quadrupole

(MQ).

We first look into the photonic crystal part without the hBN layer. Figure 1c shows the band dispersion diagram calculated near $\Gamma$ point along $Y$-$\Gamma$-$X$ without any external magnetic field, i.e., $\delta = 0$. According to Figure 1d, the simulated quality factor Q at $\Gamma$ point is greater than $10^8$, indicating that a BIC mode has been confirmed to exist. Figure 1e shows the field distribution of this BIC mode. It can be seen that the induced magnetic field of the BIC mode is distributed along the xOy plane at the four corners of the photonic crystal, which conforms to the distribution characteristics of magnetic quadrupoles. Moreover, the multipole decomposition results in Figure 1f also prove that this BIC mode in the $C_{4v}$ structure is dominated by magnetic quadrupoles. (details in Supporting Information).

When an external magnetic field is applied to this photonic crystal, the Stokes parameters under three magnetic field conditions were calculated respectively, and the far-field polarization diagrams are shown in Figure 2. The sign and magnitude of the ellipticity are determined by $S_3$ parameters, which describe the distribution of polarization states in momentum space. It can be seen that the amplitude is overall stable near the $\Gamma$ point. When $\delta = 0$, the far-field polarization diagram is completely linearly polarized near $\Gamma$ point. When $\delta = 0.033$, the far-field polarization is entirely occupied by left-handed circular polarization. Conversely, it all becomes right-handed circular polarization when $\delta = -0.033$. This demonstrates that this structure exhibits distinct chirality under different magnetic field orientations.

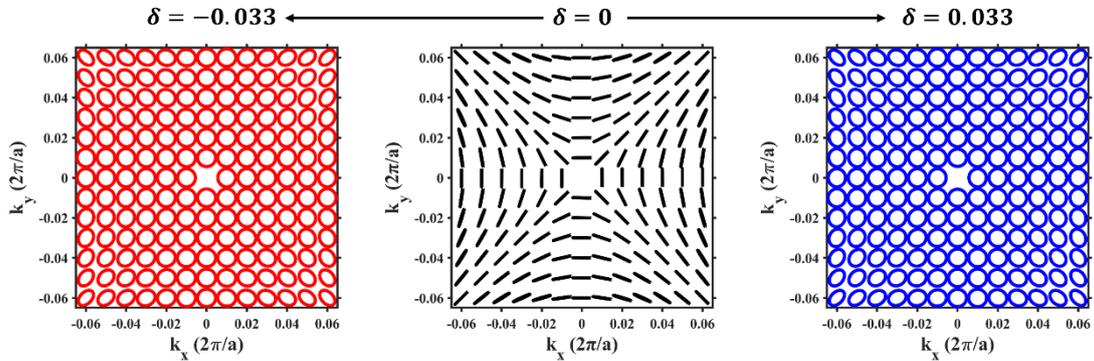

**Figure 2. Far-field polarization changes of magneto-optical photonic crystals**

**under different external magnetic field conditions.** The polarization direction is distinguished by color. Blue indicates left-handed circular polarization, red indicates right-handed circular polarization, and black indicates linear polarization.

To introduce the BIC properties into the phonon polaritonic systems, we deposited a thin layer of hBN on the surface of the magneto-optical photonic crystal. The thickness of the hBN thin layer is $t_{hBN} = 15\ nm$. hBN is a hyperbolic material and its intrinsic optical response is not affected by magnetic fields, The relative dielectric constants of hBN thin layers obtained according to the Lorentz model are as follows[46,47]:

$$\varepsilon_\alpha(\omega) = \varepsilon_{\alpha,\infty}[1 + \frac{(\omega_{LO}^\alpha)^2 - (\omega_{TO}^\alpha)^2}{(\omega_{TO}^\alpha)^2 - \omega^2 - i\gamma_\alpha\omega}] \quad (2)$$

The specific parameter values of the parallel component and the vertical component in the formula can be found in the Supporting Information (Note 2). We choose the mid-infrared band of hBN $\omega_{TO} = 1368\ \text{cm}^{-1}$, where $\omega_{TO}$ is the transverse optical phonon resonance frequency of hBN and the corresponding wavelength is $\lambda = 7310\ nm$.

In order to achieve strong coupling between the photonic crystal and the hBN phonons, we break the in-plane symmetry of the photonic crystal through the asymmetric parameter $d$, thereby transforming the BIC into a quasi-BIC mode. To investigate the effect of parameter optimization on the strong coupling of phonon polaritons, the Rabi splitting diagram varying with thickness $h$ is calculated as shown in Figure 3a. It is observed that there are two obvious upper and lower polarization branches around the phonon wavelength $\lambda = 7310\ nm$. As the thickness $h$ of the photonic crystal keeps increasing, the absorption rate of the upper polarization branch gradually decreases while that of the lower polarization branch gradually increases. It is indicated that the quasi-BIC mode caused by the asymmetric parameter $d$ may has successfully achieved strong coupling with the hBN phonon, of which Rabi splitting value is $\hbar\Omega = 5.5\ meV$. According to the temporal coupled-mode theory (TCMT)[48], the Rabi splitting value can be expressed as $\hbar\Omega = E_{UB} - E_{LB} = 2\sqrt{g^2 - \frac{(w_{phon} - w_{phot})^2}{4}}$, where $g$ is the coupling coefficient, and $w_{phon}$ and $w_{phot}$ represent the half-widths of the transmission spectrum of the hBN phonon and quasi-BIC, respectively. As can be seen from Figure S2 in the Supporting Information (Note 3), $w_{phon} = 2.3\ meV$ and $w_{phot} = 1.9\ meV$. Therefore, the coupling coefficient $g$ can be calculated from the above formula as $2.76\ meV$. It satisfies the strong coupling condition of $g > \frac{|w_{phon} - w_{phot}|}{2}$ and $g > \sqrt{\frac{w_{phon}^2 + w_{phot}^2}{2}}$, confirming that hBN and quasi-BIC have formed phonon polaritons. More details on the regulation between photon and hBN phonon coupling by geometric parameters refer to the Note 3 in the Supporting Information.

Then we traced the wavelength value at the $\Gamma$ point and obtained the Hopfield coefficients for the phonons by calculating the eigenstates of the polariton analytical model, thereby determining the fraction of phonons in the upper and lower polarization branches of the hBN polaritons (Figure 3b). It can be seen that the amplitude of the external magnetic field can adjust the ratio of phonons to photons in the polariton through the chirality of the BIC. As the absolute value of the magnetic field $|\delta|$ increases, the proportion of phonons in the upper polarization branch gradually increases, while the proportion of phonons in the lower polarization branch gradually decreases. When $\delta = 0$, the proportion of phonons in the upper polarization branch is 0.26. As $|\delta|$ increases to 0.3, the proportion of phonons in the upper branch increases to 0.35. On the contrary, in the lower polarization branch, the phonon proportion is 0.74 at $\delta = 0$. It decreases as $|\delta|$ increases and finally reaches 0.65. As can be seen the proportion of phonons in the lower polarization branch is higher than that in the upper polarization branch. These results indicate that phonon polaritons are tunable with respect to the magnetic field.

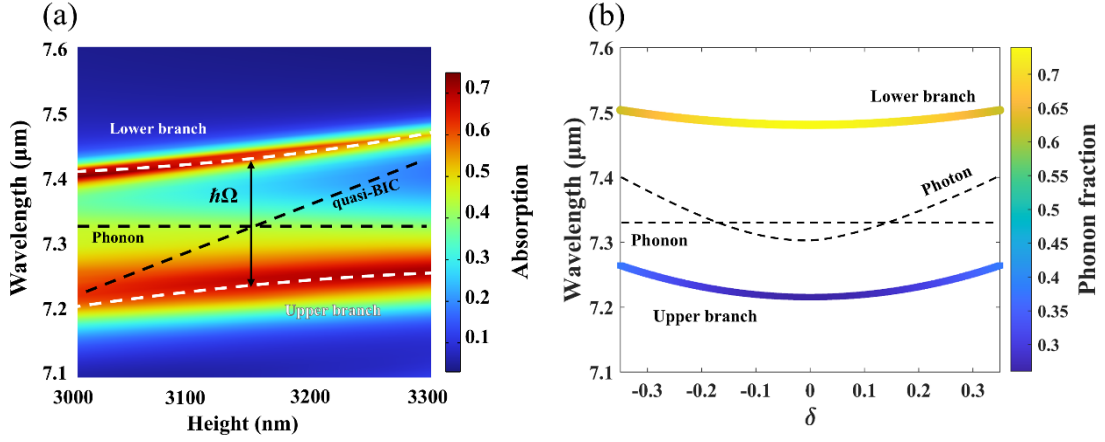

**Figure 3. Rabi splitting and phonon fraction in the phonon polaritons.** (a) The absorption contour map of the new hybrid state with different thicknesses $h$ and $d = 600\ nm$, where the black dashed lines separately represent the individual hBN phonon and magnetic quadrupole quasi-BICs mode; white dashed lines are the simulated results of the upper branch (UB) and lower branch (LB), respectively. The UB refers to the branch with the higher eigenenergy, while the LB refers to the one with the lower intrinsic energy. (b) Diagram of the variation of hBN phonon fraction in polaritons with magnetic field $\delta$.

To calculate the far-field polarization of phonon polaritons, we first constructed the effective Hamiltonian of the BIC mode. Since the structure has only one BIC energy band within the spectral range, its effective Hamiltonian can be represented by a one-dimensional irreducible representation in the $C_{4v}$ point group[49,50]. Moreover, due to the fully symmetric property of this structure, the scalar term of the effective Hamiltonian can be calculated as $\omega_0 - i\Gamma + \beta \boldsymbol{k}^2 + \gamma \boldsymbol{k}^4$ based on the projection operator of the point group representation. The specific derivation process can be found in the Note 4 of the Supporting Information. In order to calculate the far-field polarization characteristics, a 2×2 pseudo-spin splitting term $\boldsymbol{p} \cdot \boldsymbol{\sigma}$ is added to the Hamiltonian model, which is only used to represent the polarization information on the Poincaré sphere. According to the direct product relationship and the connection between the Pauli matrix and the irreducible expression of $C_{4v}$[41,51-53], the $\boldsymbol{p}$ matrices are obtained as $p_x(k) = A(2k_x k_y) + B(2k_x k_y)k^2$, $p_y(k) = C\delta$ and $p_z(k) = D(k_x^2 - k_y^2) +$

$E(k_x^2 - k_y^2)k^2$ respectively. Therefore, the final effective Hamiltonian model of this BIC mode is as follows:

$$H_{eff} = \begin{pmatrix} \Omega + p_z & d_x - ip_y \\ d_x + ip_y & \Omega - p_z \end{pmatrix} \tag{3}$$

In the formula, $\Omega = \omega_0 - i\Gamma + \beta\boldsymbol{k}^2 + \gamma\boldsymbol{k}^4$ represents the scalar term that describes the band dispersion relation.

Next, based on the effective Hamiltonian model of the photonic crystal structure, the hBN phonon term $\omega_{ph}$ was added to construct the 3×3 effective Hamiltonian of polaritons.

$$H_{pol}(\boldsymbol{k}) = \begin{pmatrix} H_{eff}^{(2)}(\boldsymbol{k},\delta) & gv(\boldsymbol{k},\delta) \\ gv^\dagger(\boldsymbol{k},\delta) & \omega_{ph} \end{pmatrix} \tag{4}$$

In the formula, $g$ represents the coupling coefficient, $\omega_{ph}$ represents the hBN phonon term, Jones vector $v(k,\delta) = (c_x, c_y)^T$ and according to the projection operator $P(k,\delta) = vv^\dagger = \frac{1}{2}(I_2 + \hat{p} \cdot \sigma)$. Based on the expectation value formula of the Pauli matrix $\langle \sigma_i \rangle = \frac{\langle v|\sigma_i|v \rangle}{\langle v|v \rangle}$ and its relationship with the eigenvalues of the Hamiltonian $H_{eff}$, we obtain $\langle \sigma_x \rangle = s_2 = \frac{p_x}{|\boldsymbol{p}|}$, $\langle \sigma_y \rangle = s_3 = \frac{p_y}{|\boldsymbol{p}|}$, $\langle \sigma_z \rangle = s_1 = \frac{p_z}{|\boldsymbol{p}|}$. Therefore, the Stokes parameters can be calculated based on the values of $\boldsymbol{p}$. The detailed derivation process and the construction of the polarization ellipse can be found in the Supporting Information (Note 4).

The eigenvectors were calculated to obtain the Stokes parameters according to the analytical method, and the far-field polarization diagrams of the polaritons under different magnetic field conditions are shown in Figure 4. Since only photons contribute to the far-field polarization, the response of polaritons to the magnetic field is consistent with that of the BIC mode shown in Figure 2. Moreover, for the upper polarization branch, the photon component is larger, so its polarization component in the $k$-space is greater than that of the lower polarization branch.

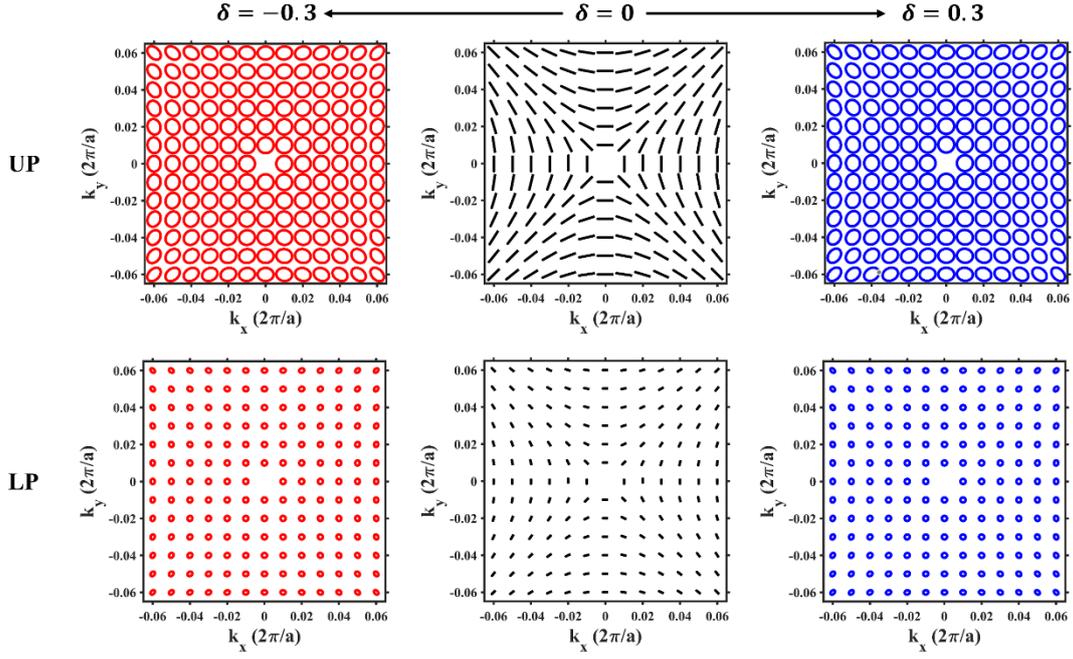

**Figure 4. Far-field polarization diagrams of hBN phonon polaritons under different magnetic field conditions.** Blue indicates left-handed circular polarization, red indicates right-handed circular polarization, and black indicates linear polarization. The size of the polarization ellipse and the length of the polarization line are related to the photon proportion calculated from the eigenvalues.

In order to further verify the selective response of hBN phonon polaritons to magnetic fields, we selected a set of geometric parameters $a = 4100\ nm, L = 3050\ nm, r = 600\ nm, h = 3000\ nm, d = 600\ nm,$ and applied different longitudinal magnetic fields to the hBN-photonic crystal heterostructure. Meanwhile, left-hand circularly polarized light and right-hand circularly polarized light were incident on them, respectively. The absorption spectra of the phonon polaritons obtained are shown in Figure 5 by using the method of COMSOL Multiphysics simulation.

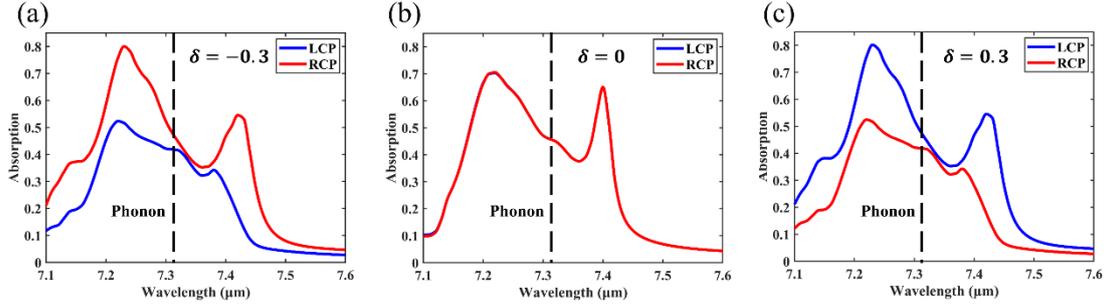

**Figure 5. Absorption spectra of the hBN phonon polariton under left-hand circular polarization and right-hand circular polarization incidences.** (a) Magnetic field with $\delta = -0.3$, (b) no magnetic field and (c) magnetic field with $\delta = 0.3$ are applied to the magneto-optical photonic crystal.

First of all, it is clearly observable from the three absorption spectra shown in Figure 5 that two peaks split around the wavelength of hBN phonon. Moreover, it can be clearly seen that when there is no magnetic field applied ($\delta = 0$), the absorption peaks of hBN phonon polaritons are the same regardless of whether left-hand circularly polarized light is incident or right-hand circularly polarized light is incident from the identical spectra in Figure 5b. However, the response of polaritons to left-hand circularly polarized light is apparently higher than that to right-hand circularly polarized light when $\delta = 0.3$. Similarly, the polaritons are more sensitive to the excitation of right-hand circularly polarized light when $\delta = -0.3$. Since absorptions of the polaritons all come from the phonon part, the results in this Figure directly indicate that the chirality of the magnetic quadrupole BIC is transferred to the polariton through the hBN phonon.

Based on Figure 3 to 5, since the polariton is a hybrid mode formed by coupling a quasi-BIC photon to an hBN phonon, the hBN phonon is essentially an achiral mode that couples with the fully chiral BIC, the net chirality of the mixed state will be weakened, leading to only partial circular dichroism of the hBN polariton under the influence of the magnetic field. This clarifies the reason why this phonon polariton cannot exhibit perfect absorption of the LCP or RCP incident lights. Therefore, as shown in Figure 5c, even when a positive magnetic field is applied, the polariton can still respond to the right-hand circularly polarized light. This part of the response is caused by the hBN phonon. The same applies to the negative magnetic field shown in Figure 5a. These results indicate that the observed polaritonic chirality originates from the BIC component rather than from the hBN phonon, which means chirality was transferred to the hBN polariton through the BIC mode.

## 3. Conclusions

In conclusion, we have proposed and analyzed a magnetically tunable chiral phonon polariton platform that consists of a thin hBN layer strongly coupled with a magneto-optical photonic crystal. We first designed and theoretically studied the underlying magneto-optical BIC photonic crystal using the effective Hamiltonian and far-field polarization analysis, which exhibits chiral characteristics with respect to the magnetic field direction. Then we investigated the properties of phonon polaritons and revealed that the magnetic field could effectively alter the proportion of photons and phonons in the phonon polaritons, enabling magnetic tunability of phonon polaritons in the system. We also demonstrated that the chirality of the magnetic-responsive BIC was successfully transferred to the polariton through the hBN phonon strong coupling by absorption spectra and far-field polarization. In particular, the absorption spectra show handedness-selective response depending on the direction of the magnetic field, enabling possible applications in magneto-optical sensors. In this magnetic-responsive and tunable chiral phonon polaritonic platform, the magnetic field provides a unique, external, reversible, and continuously adjustable degree of freedom for chiral phonon polaritons that is distinct from existing routes, leading to magnetic tunable light-matter interaction. This enables a rich platform for studying the spin, valley, and chirality interactions in light-matter coupling, and further extends to possible applications such as non-reciprocal integrated photonics, reconfigurable circularly polarized luminescence, and active chirality sensing.

**References**

(1)  Sanvitto, D.; Kéna-Cohen, S. The Road towards Polaritonic Devices. Nat. Mater. 2016, 15, 1061-1073.

(2)  Amo, A.; Lefrère, J.; Pigeon, S.; et al. Superfluidity of Polaritons in Semiconductor Microcavities. Nat. Phys. 2009, 5, 805-810.

(3)  Ballarini, D.; De Giorgi, M.; Cancellieri, E.; et al. All-Optical Polariton Transistor. Nat. Commun. 2013, 4, 1778.

(4)  Kasprzak, J.; Richard, M.; Kundermann, S.; et al. Bose-Einstein Condensation of Exciton Polaritons. Nature 2006, 443, 409-414.

(5)  Amo, A.; Liew, T.; Adrados, C.; et al. Exciton–Polariton Spin Switches. Nat. Photonics 2010, 4, 361-366.

(6)  Lerario, G.; Fieramosca, A.; Barachati, F.; et al. Room-Temperature Superfluidity in a Polariton Condensate. Nat. Phys. 2017, 13, 837-841.

(7)  Li, M.; Sinev, I.; Benimetskiy, F.; et al. Experimental Observation of Topological Z2 Exciton-Polaritons in Transition Metal Dichalcogenide Monolayers. Nat. Commun. 2021, 12, 4425.

(8)  Teng, H.; Jiang, C.; Liu, M.; Qu, Y.; Zhou, S.; Xue, Z.; Zhu, H.; Gui, J.; Xi, S.; Yang, Y.; Chen, N.; Hu, H.; Dai, Q. Programmable Re-entrant Topological Polaritons in Graphene Grating/α-MoO3 Heterostructure. Nano Lett. 2026, 26 (8), 2923-2931.

(9)  Ma, Y.; Zhong, G.; Dai, Z.; Ou, Q. In-plane Hyperbolic Phonon Polaritons: Materials, Properties, and Nanophotonic Devices. Nanophoton. 2024, 1, 25.

(10)  Taboada-Gutiérrez, J.; Zhou, Y.; Tresguerres-Mata, A. I. F.; Lanza, C.; Martínez-Suárez, A.; Álvarez-Pérez, G.; Duan, J.; Martín, J. I.; Vélez, M.; Prieto, I.; Bercher, A.; Teyssier, J.; Errea, I.; Nikitin, A. Y.; Martín-Sánchez, J.; Kuzmenko, A. B.; Alonso-González, P. Unveiling the Mechanism of Phonon-Polariton Damping in α-MoO3. ACS Photonics 2024, 11, 3570–3577.

(11)  Xu, R.; Crassee, I.; Bechtel, H. A.; Zhou, Y.; Bercher, A.; Korosec, L.; Rischau, C. W.; Teyssier, J.; Crust, K. J.; Lee, Y.; Gilbert Corder, S. N.; Li, J.;

Dionne, J. A.; Hwang, H. Y.; Kuzmenko, A. B.; Liu, Y. Highly Confined Epsilon-Near-Zero and Surface Phonon Polaritons in SrTiO3 Membranes. Nat. Commun. 2024, 15, 4743.

(12) He, P.; Li, J.; Li, C.; Li, N.; Han, B.; Shi, R.; Qi, R.; Du, J.; Yu, P.; Gao, P. Strongly Confined Mid-Infrared to Terahertz Phonon Polaritons in Ultrathin SrTiO3. Sci. Adv. 2025, 11, eady7316.

(13) Mancini, A.; Nan, L.; Wendisch, F. J.; Berté, R.; Ren, H.; Cortés, E.; Maier, S. A. Near-Field Retrieval of the Surface Phonon Polariton Dispersion in Free-Standing Silicon Carbide Thin Films. ACS Photonics 2022, 9, 3696–3704.

(14) Hutchins, W.; Zare, S.; Hirt, D. M.; Tomko, J. A.; Matson, J. R.; Diaz-Granados, K.; Long, M., III; He, M.; Pfeifer, T.; Li, J.; Edgar, J. H.; Maria, J.-P.; Caldwell, J. D.; Hopkins, P. E. Ultrafast Evanescent Heat Transfer across Solid Interfaces via Hyperbolic Phonon–Polariton Modes in Hexagonal Boron Nitride. Nat. Mater. 2025, 24, 698–706.

(15) Yang, J.; Tang, J.; Ghasemian, M. B.; Mayyas, M.; Yu, Q. V.; Li, L. H.; Kalantar-Zadeh, K. High-Q Phonon-Polaritons in Spatially Confined Freestanding α-MoO3. ACS Photonics 2022, 9, 905–913.

(16) Conrads, L.; Schüler, L.; Wirth, K. G.; Wuttig, M.; Taubner, T. Direct Programming of Confined Surface Phonon Polariton Resonators with the Plasmonic Phase-Change Material In3SbTe2. Nat. Commun. 2024, 15, 3472.

(17) Zhou, Y.; Waelchli, A.; Boselli, M.; Crassee, I.; Bercher, A.; Luo, W.; Duan, J.; van Mechelen, J. L. M.; van der Marel, D.; Teyssier, J.; Rischau, C. W.; Korosec, L.; Gariglio, S.; Triscone, J.-M.; Kuzmenko, A. B. Thermal and Electrostatic Tuning of Surface Phonon-Polaritons in LaAlO3/SrTiO3 Heterostructures. Nat. Commun. 2023, 14, 7686.

(18) Noh, B.-I.; Jafari Ghalekohneh, S.; Chen, M.; Shen, J.; Janzen, E.; Zhou, L.; Chen, P.; Edgar, J. H.; Zhao, B.; Dai, S. Mode Conversion of Hyperbolic Phonon Polaritons in van der Waals Terraces. Nat. Commun. 2026, 17, 1273.

(19) Li, D.; Pan, Z.; Caldwell, J. D. Phonon Polariton-Mediated Heat Conduction: Perspectives from Recent Progress. J. Mater. Res. 2024, 39,

3193–3201.

(20) Pei, Y.; Chen, L.; Jeon, W.; Liu, Z.; Chen, R. Low-Dimensional Heat Conduction in Surface Phonon Polariton Waveguide. Nat. Commun. 2023, 14, 8242.

(21) Fang, K.; Yu, Z.; Liu, V.; Fan, S. Ultracompact nonreciprocal optical isolator based on guided resonance in a magneto-optical photonic crystal slab. Opt. Lett. 2011, 36 (21), 4254–4256.

(22) Wang, Z.; Fan, S. Optical circulators in two-dimensional magneto-optical photonic crystals. Opt. Lett. 2005, 30 (15), 1989–1991.

(23) Król, M.; Mirek, R.; Lekenta, K.; Rousset, J.-G.; Stephan, D.; Nawrocki, M.; Matuszewski, M.; Szczytko, J.; Pacuski, W.; Piętka, B. Spin polarized semimagnetic exciton-polariton condensate in magnetic field. Sci. Rep. 2018, 8, 6694.

(24) Wang, T.; Zhang, D.; Yang, S.; Lin, Z.; Chen, Q.; Yang, J.; Gong, Q.; Chen, Z.; Ye, Y.; Liu, W. Magnetically-dressed CrSBr exciton-polaritons in ultrastrong coupling regime. Nat. Commun. 2023, 14, 5966.

(25) Nessi, L.; Occhialini, C. A.; Demir, A. K.; Powalla, L.; Comin, R. Magnetic field tunable polaritons in the ultrastrong coupling regime in CrSBr. ACS Nano 2024, 18 (50), 34235–34243.

(26) Zhen, B.; Hsu, C. W.; Lu, L.; Stone, A. D.; Soljačić, M. Topological Nature of Optical Bound States in the Continuum. Phys. Rev. Lett. 2014, 113, 257401.

(27) Hsu, C. W.; Zhen, B.; Lee, J.; et al. Observation of Trapped Light within the Radiation Continuum. Nature 2013, 499, 188-191.

(28) Jin, J.; Yin, X.; Ni, L.; et al. Topologically Enabled Ultrahigh-Q Guided Resonances Robust to Out-of-Plane Scattering. Nature 2019, 574, 501-504.

(29) Kang, M.; Liu, T.; Chan, C. T.; et al. Applications of Bound States in the Continuum in Photonics. Nat. Rev. Phys. 2023, 5, 659-678.

(30) Kuznetsov, A. I.; et al. Optically Resonant Dielectric Nanostructures. Science 2016, 354, aag2472.

(31) Kawasaki, D.; Tanaka, T. High-Q Metasurface Absorber Enabled by Symmetry Breaking in a Plasmonic Lattice. Nano Lett. 2026, 26 (9), 3195-3203.

(32) Wang, W.; Srivastava, Y. K.; Tan, T. C.; Wang, Z.; Singh, R. Brillouin zone folding driven bound states in the continuum. Nat. Commun. 2023, 14, 2811.

(33) Li, X.; Ma, J.; Liu, S.; Huang, P.; Chen, B.; Wei, D.; Liu, J. Efficient second harmonic generation by harnessing bound states in the continuum in semi-nonlinear etchless lithium niobate waveguides. Light Sci. Appl. 2022, 11, 317.

(34) Qin, H.; Chen, S.; Zhang, W.; Zhang, H.; Pan, R.; Li, J.; Shi, L.; Zi, J.; Zhang, X. Optical moiré bound states in the continuum. Nat. Commun. 2024, 15, 9080.

(35) Luo, W.; Zhao, R.; Liu, Y.; et al. Desymmetrized Metamaterials Enable Perfect Absorption. Adv. Mater. 2026, e22888.

(36) Zang, M. Y.; Zhang, J. S.; Lou, Q.; Zhou, S.; Fu, S. F.; Zhang, Q.; Wang, X. G.; Wang, X. Z. Terahertz Sensor Based on Bound States in the Continuum of an h-BN Metasurface. J. Opt. Soc. Am. B 2025, 42 (9), 2032–2039.

(37) Nan, L.; Mancini, A.; Weber, T.; Seah, G. L.; Cortés, E.; Tittl, A.; Maier, S. A. Angular Dispersion Suppression in Deeply Subwavelength Phonon Polariton Bound States in the Continuum Metasurfaces. Nat. Photonics 2025, 19, 615–623.

(38) Gupta, H.; Venturi, G.; Contino, T.; Janzen, E.; Edgar, J. H.; De Angelis, F.; Toma, A.; Ambrosio, A.; Tamagnone, M. Bound States in the Continuum and Long-Range Coupling of Polaritons in Hexagonal Boron Nitride Nanoresonators. ACS Photonics 2024, 11 (10), 4017–4026.

(39) Sun, M. Z.; Zhang, Q.; Jin, P.; Zhu, Y. H.; Fu, S. F.; Zhang, Q.; Zhou, S.; Wang, X. G.; Wang, X. Z. High Quality Factor of Bound States in Continuum in hBN Metasurface. J. Appl. Phys. 2024, 136 (13), 133103.

(40) Yang, J.; Zhang, L.; Wang, K.; Zhang, C.; Fan, A.; He, Z.; Li, Z.; Han, X.; Ling, F.; Lu, P. Manipulating Terahertz Phonon-Polariton in the Ultrastrong

Coupling Regime with Bound States in the Continuum. Light Sci. Appl. 2025, 14, 360.

(41) Zhao, X.; Wang, J.; Liu, W.; Che, Z.; Wang, X.; Chan, C. T.; Shi, L.; Zi, J. Spin-Orbit-Locking Chiral Bound States in the Continuum. Phys. Rev. Lett. 2024, 133 (3), 036201.

(42) Lv, W.; et al. Robust Generation of Intrinsic C Points with Magneto-Optical Bound States in the Continuum. Sci. Adv. 2024, 10, eads0157.

(43) Zhao, X.; Li, X.; Yang, S.; Zhang, C.; Zhang, K.; Kong, W. Tailoring Polarization Singularities in Degenerate Bands through Structural Symmetry and Magnetic Field. Opt. Lett. 2025, 50 (23), 7235-7238.

(44) Armelles, G.; Cebollada, A.; García-Martín, A.; González, M. U. Magnetoplasmonics: Combining Magnetic and Plasmonic Functionalities. Adv. Opt. Mater. 2013, 1, 10-35.

(45) Haider, T. A Review of Magneto-Optic Effects and Its Application. Int. J. Electromagn. Appl. 2017, 7, 17.

(46) Guddala, S.; et al. Topological Phonon-Polariton Funneling in Midinfrared Metasurfaces. Science 2021, 374, 225-227.

(47) Yang, J.; Krix, Z. E.; Kim, S.; Tang, J.; Mayyas, M.; Wang, Y.; Watanabe, K.; Taniguchi, T.; Li, L. H.; Hamilton, A. R.; Aharonovich, I.; Sushkov, O. P.; Kalantar-Zadeh, K. Near-Field Excited Archimedean-like Tiling Patterns in Phonon-Polaritonic Crystals. ACS Nano 2021, 15 (5), 9134-9142.

(48) Liu, X.; Galfsky, T.; Sun, Z.; et al. Strong Light–Matter Coupling in Two-Dimensional Atomic Crystals. Nat. Photonics 2015, 9, 30-34.

(49) Luttinger, J. M.; Kohn, W. Motion of Electrons and Holes in Perturbed Periodic Fields. Phys. Rev. 1955, 97, 869-883.

(50) Minkov, M.; Williamson, I. A. D.; Xiao, M.; Fan, S. Zero-Index Bound States in the Continuum. Phys. Rev. Lett. 2018, 121, 263901.

(51) Bir, G. L.; Pikus, G. E. Symmetry and Strain-Induced Effects in Semiconductors; Wiley: New York, 1974.

(52) Samokhin, K. V. On the Effective Models of Spin–Orbit Coupling in a

Two-Dimensional Electron Gas. Ann. Phys. 2022, 437, 168710.

(53)   Chen, A.; Liu, W.; Zhang, Y.; Wang, B.; Liu, X.; Shi, L.; Lu, L.; Zi, J. Observing Vortex Polarization Singularities at Optical Band Degeneracies. Phys. Rev. B 2019, 99, 180101.

**Acknowledgements**

We thank financial support from the National Key R&D Program of China (Grant No. 2023YFB3811400), Shenzhen Science and Technology Innovation Commission (Grant No.20231204145504001), National Natural Science Foundation of China (Grant No.12404435), and Guangdong Science and Technology Commission (Grant No.2025A1515010774).

**Author information**

All authors contributed to the work substantially.

**Conflict of interest**

The authors declare no competing interests.